\documentclass{iau_FM}
\usepackage{graphicx}
\graphicspath{{./}{figures/}}
\usepackage{hyperref}
\hypersetup{urlcolor=blue, colorlinks=true, linkcolor=blue, citecolor=blue}

\usepackage{natbib}

\newcommand{\astronomaly}{\textsc{astronomaly}}
\newcommand{\protege}{\textsc{astronomaly: protege}}

\title[FM 7.~~Enabling New Discoveries with Machine Learning] 
{Enabling New Discoveries with \\Machine Learning}

\author[Michelle Lochner]   
{Michelle Lochner$^{1,2}$}

\affiliation{$^1$Department of Physics and Astronomy, University of the Western Cape, \\Bellville, Cape Town, 7535, South Africa \\ email: {\tt mlochner@uwc.ac.za} \\[\affilskip]
$^2$South African Radio Astronomy Observatory, Liesbeek House, River Park, \\Liesbeek Parkway, Mowbray 7705, South Africa}

\pubyear{2024}
\jname{Astronomy in Focus, Focus Meeting 7} 
\editors{Shazrene Mohamed, Russ Taylor and Bhargav Vaidya, ed.}
\begin{document}

\maketitle

\begin{abstract}
The next generation of telescopes such as the Square Kilometre Array and the Vera C. Rubin Observatory will produce enormous quantities of data, too large for traditional analysis techniques. Machine learning has proven invaluable in handling massive data volumes and automating tasks traditionally done by human scientists. The question is, can machine learning also automate scientific discovery? Recent advances in the field have made this feasible but only when combined with expert human input through active learning. \astronomaly{} is a publicly available framework for active anomaly detection in astronomy. It learns to make recommendations of interesting sources to a user, allowing scientists to quickly uncover rare and new classes of objects in large datasets. In this paper, we review some of the discoveries made using \astronomaly{}, discuss the latest deep learning-based approaches and explore the critical role of the cyber-human interface in unveiling cosmic mysteries hidden in massive astronomical datasets.
\end{abstract}

\firstsection 
\section{Introduction}
In 1967, Dame Jocelyn Bell Burnell made one of the most important discoveries in astronomy history \citep{Hewish1968}. What it took to discover the first pulsar was a bright graduate student, carefully looking through all her data and noticing something odd. The Square Kilometre Array, under construction in South Africa and Australia, will discover over a billion radio galaxies and thousands of pulsars and fast radio transients \citep{Braun2015}. When the ten-year Legacy Survey of Space and Time on the Vera C. Rubin Observatory (being built in Chile) is complete, its catalogue of optical galaxies will number roughly 20 billion \citep{Ivezic2019}. Equally incredible will be the alert rate of the Rubin Observatory: its large field of view and sensitivity means it will discover something changing on the sky over 10 million times a night \citep{Ivezic2019}. The question is, how do we make incredible, unexpected discoveries, like that of the pulsar, when it will simply be impossible to visually inspect all the data?

Machine learning has had a remarkable impact on essentially every aspect of human life, from interactive language models \citep{Brown2020} to video recommendations \citep{Roy2022}, and astronomy is no exception. These tools also have the potential to automate scientific discovery. Anomaly detection, the automatic identification of novel and outlier sources, generally falls under a branch of machine learning known as unsupervised learning. These algorithms do not require labels (usually obtained from humans) for training and can operate on the data directly. This is critical as, by definition, one cannot have labels for rare or new objects that have never been seen before. However, just because something is anomalous, does not mean it is scientifically interesting.

\section{Astronomaly}
Many objects in a dataset, such as artefacts, will appear anomalous to a machine learning algorithm but will not be scientifically interesting. However, the algorithm can be improved by providing a small amount of strategically chosen training data, in a process known as active learning. Active learning aims to optimally combine the raw processing power of machine learning with the knowledge and intuition of a human scientist, with the goal being to minimise the amount of data the human has to sort through.

\citet{Lochner2021} introduced \astronomaly{}, a general framework for anomaly detection in astronomy. It includes a novel approach to active learning and combines a Python back end with a JavaScript front end to obtain scores of ``interestingness'' from a human user. \astronomaly{} can rapidly sort through a dataset, prioritising sources likely to be of interest to the user and relegating boring sources to the bottom of the pile.

\begin{figure}
\begin{center}
 \includegraphics[width=\linewidth]{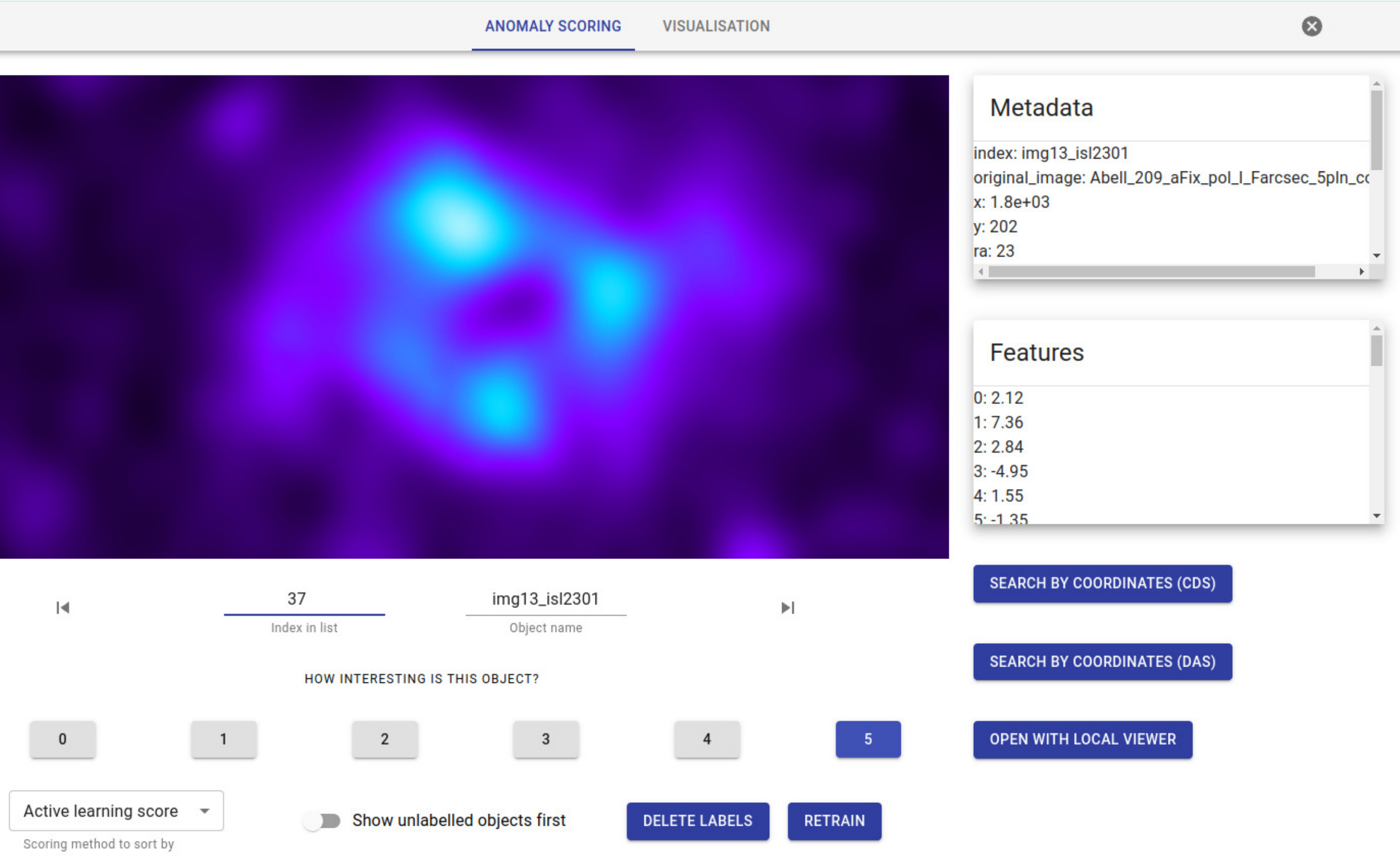}
 \caption{Screenshot of the \astronomaly{} front end. The source featured is SAURON, which was first discovered in radio data using \astronomaly{} \citep{Lochner2023}.}
   \label{fig:screenshot}
\end{center}
\end{figure}

\astronomaly{} can be used for almost any type of data. \cite{Lochner2021} showed that it could detect unusual sources such as mergers in the optical images of the citizen science Galaxy Zoo \citep{Lintott2008, Lintott2011} dataset. \citet{Webb2020} used \astronomaly{} to discover new variable and flare stars from among  85 553 optical light curves. Finally, \astronomaly{} demonstrated its capabilities by making a genuine discovery of a new type of radio source called ``SAURON'' (a Steep and Uneven Ring of Nonthermal radiation), the physics of which is still undetermined \citep{Lochner2023}. This source, discovered in radio data from the MeerKAT Galaxy Cluster Legacy Survey \citep{Knowles2022}, does not resemble any known source and could be the remnant of a supermassive black hole merger. \autoref{fig:screenshot} shows a screenshot of \astronomaly{} highlighting SAURON.

\section{Feature Extraction}
\begin{figure}
\begin{center}
 \includegraphics[width=\linewidth]{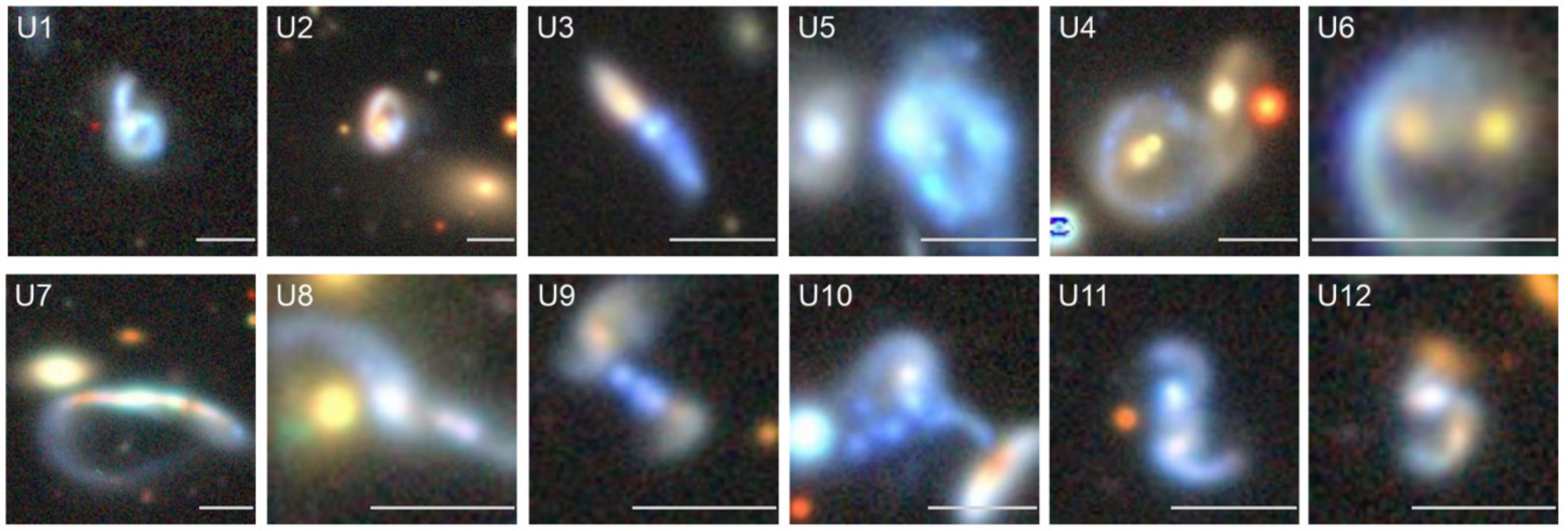}
 \caption{The twelve optical sources from the DECaLS survey rated the most interesting by \astronomaly{}, adapted from \citet{Etsebeth2024}. The scale bar in each image represents 10 arcseconds.}
   \label{fig:etsebeth_anomalies}
\end{center}
\end{figure}
Machine learning algorithms in general cannot work directly with raw data, which is usually complex and high-dimensional. Instead, feature extraction must be used to find an informative, low-dimensional representation of the original data. This is traditionally the most difficult step in any machine learning process, as it can require significant domain knowledge to determine what features may be sufficiently informative for the algorithm. Yet it is also the most important step, as the features selected will ultimately dictate what type of anomalies the algorithm will be sensitive to. 

For example, \citet{Lochner2021} used hand-crafted features which captured the morphology of optical galaxies (by fitting ellipses to isophotes and using the parameters as features), but these features would have been completely insensitive to galaxies with unusual colours. \citet{Webb2020} used a domain-specific set of features in applying anomaly detection to light curves. While this worked well in detecting variable and flare stars, these features may be less sensitive to other types of anomalies.

In the last few decades, deep learning has revolutionised the field of supervised learning primarily because of its ability to learn relevant features automatically from the data. Deep neural networks have successive layers of neurons which learn features on multiple scales which best solve the task the network is training for. Although these tools have been developed primarily for supervised learning, where labelled training data are available, we have begun making use of this machinery in the unsupervised realm, primarily to perform feature extraction before a more traditional anomaly detection algorithm is applied. 

\citet{Muthukrishna2022} applied a deep learning algorithm to time series data with the aim of detecting anomalies. This algorithm was trained, in a supervised manner, to predict the next value in a light curve and any value that deviated significantly from the prediction was flagged as anomalous. However, \citet{Muthukrishna2022} came to the surprising conclusion that deep learning was simply too flexible to be useful as it could fit essentially any source. However, by making use of a pretrained classifier as a feature extractor and a more sophisticated anomaly detection algorithm, \citet{Gupta2024} instead found that deep learning-derived features were valuable in detecting anomalous light curves. 

\citet{Walmsley2022} pioneered the approach of using a pretrained network as a feature extractor for optical Galaxy Zoo images, showing that an algorithm trained on a complex classification task creates a representation of the galaxy images which can be used for downstream tasks, including anomaly detection. We extended this approach in \citet{Etsebeth2024}, applying \astronomaly{} to nearly 4 million optical images from the Dark Energy Camera Legacy Survey (DECaLS, \citealp{Dey_2019}), and discovered the interesting sources highlighted in \autoref{fig:etsebeth_anomalies}.

\begin{figure}
\begin{center}
 \includegraphics[width=\linewidth]{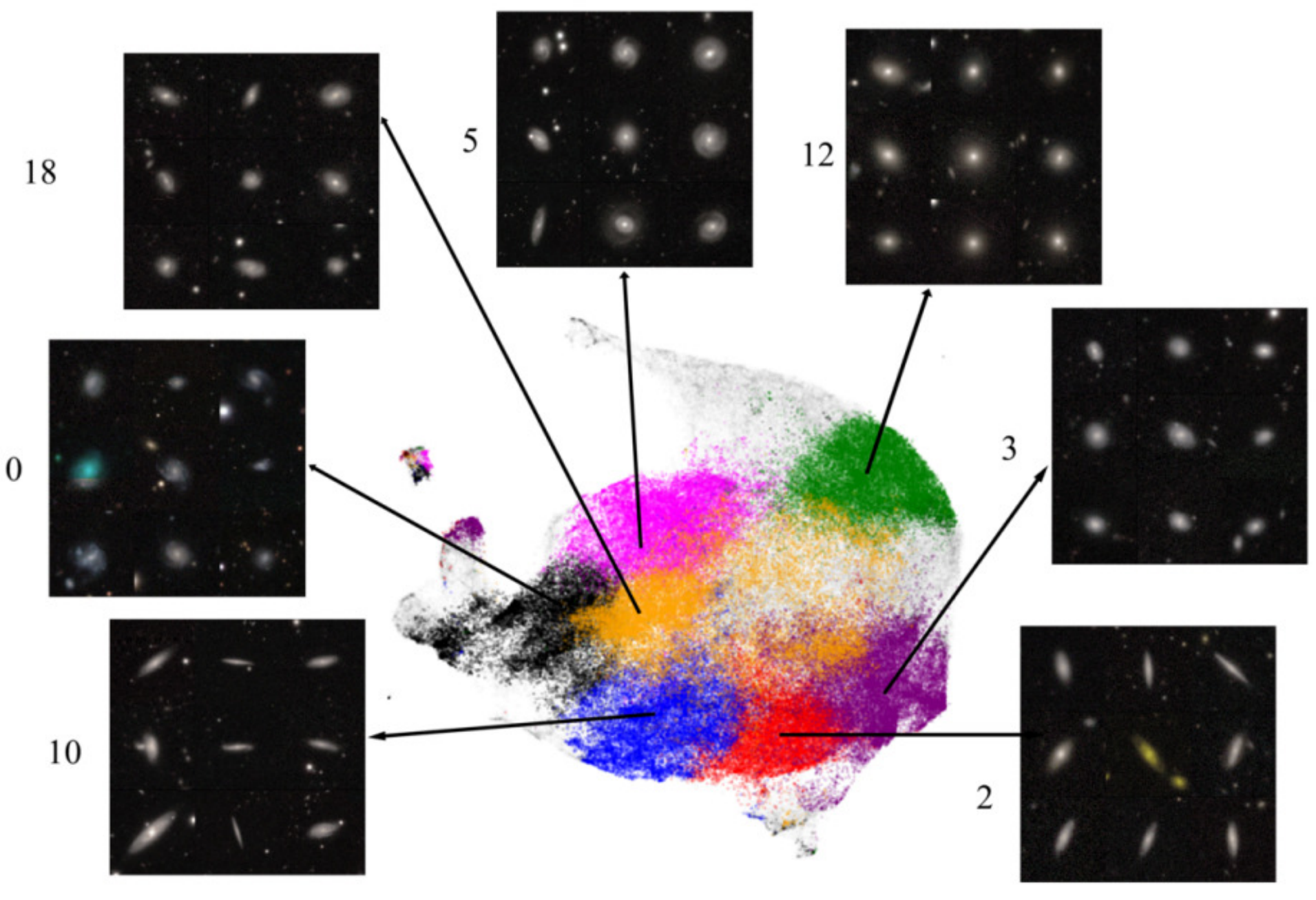}
 \caption{UMAP plot for Galaxy Zoo images with automatic clustering (highlighted by different colours) grouping together different sources \citep{Mohale2024}.}
   \label{fig:mohale_clusters}
\end{center}
\end{figure}

The problem with using a pretrained network as a feature extractor is that the data it is being applied to needs to be fairly similar to the data it was trained on. This is a challenge for fields such as radio astronomy where large, clearly labelled datasets are simply not available. In \citet{Mohale2024}, we turned to self-supervised learning to extract features for any image dataset regardless of the existence of any labels. \autoref{fig:mohale_clusters} shows that, using the representation from a self-supervised algorithm, we could automatically cluster similar looking galaxies together and subsequently perform successful anomaly detection. The central plot in this figure is known as a Uniform Manifold Approximation and Project plot \citep[UMAP,][]{McInnes2018} and is a low-dimensional embedding of the high-dimensional feature space, such that each point represents an entire image of a galaxy and the images corresponding to nearby points \emph{should} be similar, if the feature extraction process is successful.

\section{Interesting vs. Anomalous}
\begin{figure}
\begin{center}
 \includegraphics[width=\linewidth]{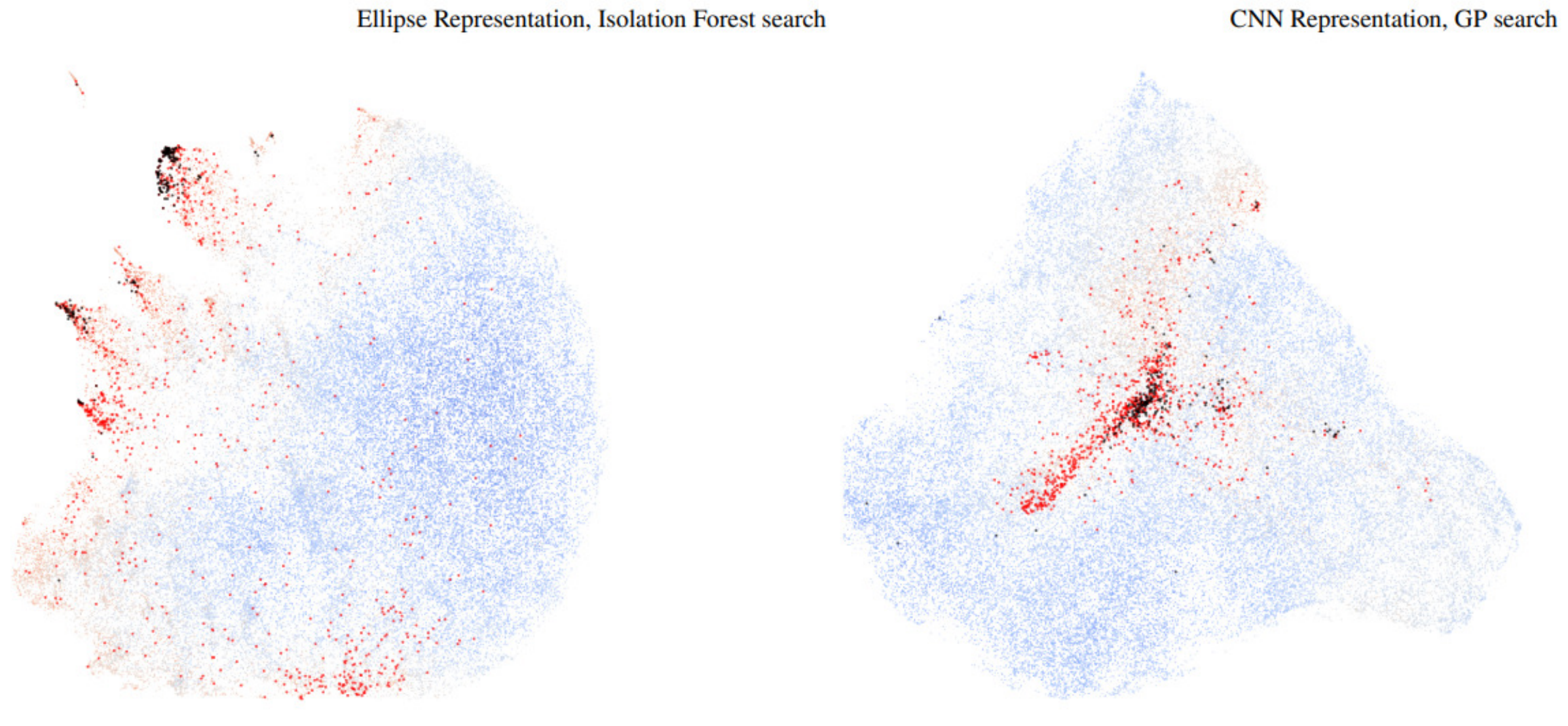}
 \caption{UMAP for hand-crafted morphological features (left) and deep learning-derived features (right) with unusual sources coloured red and black \citep{Walmsley2022}.}
   \label{fig:walmsley_umap}
\end{center}
\end{figure}
Whether from pretrained classifiers or self-supervised learning, we have found that deep learning is generally more successful at extracting meaningful features than traditional approaches. An important challenge remains though and is illustrated in \autoref{fig:walmsley_umap}. The UMAP plot on the left is for the hand-crafted morphological features described in \citet{Lochner2021}, showing that while the sources citizen scientists rated as ``odd'' (shown in red and black) are quite spread out, they predominantly appear towards the edge of feature space and thus would trigger as ``anomalous'' to a machine learning algorithm. In contrast, the plot on the right, for the deep learning-derived features, shows far better clustering of the same ``odd'' sources, but they are now deeply embedded in feature space meaning they will not be identifiable by a traditional approach. We have found this phenomenon occurs frequently when using deep learning for feature extraction.

\citet{Walmsley2022} developed a fundamentally different approach to identify these sources, which we have subsequently begun to call \protege{}. This technique bypasses any true anomaly detection and directly uses active learning to learn what is ``interesting'' according to the user via a small amount of iterative data labelling. This is philosophically different than anomaly detection but, given that active learning is required anyway to remove uninteresting (according to the user) sources, it is able to be used to achieve the same goals of scientific discovery. Indeed, we find coupling \protege{} with self-supervised deep learning to be immensely powerful in making discoveries in large and unlabelled datasets, with minimal human effort (Lochner and Rudnick, in prep).

\section{Conclusions}
With the rapidly increasing size and richness of astronomical datasets, machine learning for anomaly detection has become a feverishly active field of research. We have shown that \astronomaly{} is highly effective at discovering new and rare classes of sources in large datasets of astronomical images and light curves. Active learning plays a critical role because anomalous is not the same as interesting. Indeed, we would argue that, since anomalies are difficult to define, it may be more natural to instead embrace the subjective nature of discovery and rely on approaches, such as \protege{}, that are effective at detecting ``interesting'' sources, as defined by human users. While \protege{} learns from a human, it can be trained to learn a somewhat more general sense of ``interesting'' and recommend sources to a user that they may not have considered themselves. This cyber-human interface leverages the best of human scientists and machine learning and is the key to enabling new discoveries in the data deluge to come.

\bibliography{refs}{}
\bibliographystyle{iaulike}

\end{document}